\documentclass[preprint,12pt,3p]{elsarticle}

\usepackage{amssymb}
\usepackage{graphicx}
\usepackage[cmex10]{amsmath}
\usepackage{array}
\usepackage{mdwmath}
\usepackage{caption}
\usepackage{multirow}
\usepackage[linesnumbered,ruled,vlined]{algorithm2e}
\SetKwInput{KwInput}{Input}                
\SetKwInput{KwOutput}{Output}              

\journal{Journal version of the AINA-24 conference paper}

\begin{document}

\begin{frontmatter}
	
	\title{Next-generation optical networks to sustain connectivity of the future: All roads lead to optical-computing-enabled network?}

	\author[label1]{Dao Thanh Hai \corref{cor1}}
	\address[label1]{School of Science, Engineering and Technology, RMIT University Vietnam}
	
	
	\ead{hai.dao5@rmit.edu.vn}
	
	\author[label5]{Isaac Woungang}
	\address[label5]{Department of Computer Science, Toronto Metropolitan University}
	
	\ead{iwoungan@torontomu.ca}
	\begin{abstract}
		The rise and then rapid developments of various nascent technologies, encompassing notably Internet of Things (IoT), Big Data and Artificial Intelligence (AI) have been heralding a new era of connectivity, spanning from people, things, to ultimately intelligence. Such connectivity of the future will be expected to drive explosive Internet traffic growths and thus, posing unprecedented challenges for network operators in scaling up the capacity in a greater cost and energy efficiency. Optical communications and networks constituting the backbone of Internet infrastructure will thus have to be radically different in the next 10 years and beyond. Indeed, there have been a number of on-going technological innovations holding the promises of many orders-of-magnitude capacity expansion, notably multi-band and/or spatial-division-multiplexing-based technologies. On the other hand, from an architectural perspective with the main goal of reducing the effective traffic load in the network and thus gaining more operational efficiency, optical networks have been essentially remained the same in the recent two decades since the year 2000s with the success and then dominance of optical-bypass mode, featuring both significant cost and energy savings compared to the predecessor optical-electrical-optical mode, by eliminating massive intermediate optical-electrical-optical interfaces. In the optical-bypass-enabled network, the add/drop and cross-connect functions constitute the fundamental operations in handling the traffic at the optical layer, whose the underlying principle lies in the fact that in cross-connecting in-transit lightpaths over an intermediate node, such lightpaths must be guarded from each other in a certain dimension, be it the time, frequency or spatial domain, to avoid interference, which is treated as destructive. In view of the rapid progresses in the realm of optical computing enabling the controlled interference between optical channels that are tailored for various computing capabilities, we envision a different perspective to turn around the long-established wisdom in optical-bypass network by putting the optical channel interference to a good use, resulting into the so-called \textit{optical-computing-enabled network}, which potentially marks the next frontier of optical network architecture with the arrival of optical communication-computing integrated networks. The proposal of ours is essentially characterized by the new capability at optical nodes permitting the superposition of transitional lightpaths to compute new ones of better spectrum utilization and/or for special computing purposes such as large-scale AI training while ensuring the traffic accommodations. In underlining the potential merits of bringing in-network optical computing functions into the optical layer, this paper presents two illustrative examples based on the optical aggregation and optical XOR operations which have been progressively maturing and thus, could be feasibly integrated into the current legacy infrastructure with minimal disruptions. As a departure from optical-bypass operation, the new optical computing capabilities available at the optical nodes thus imply a radical change in the network design problems and deriving the associated algorithmic solutions, which are broadly termed as optical network design and planning 2.0, so that the capital and operational efficiency could be fully unlocked. As a proof-of-concept of this paradigm shift, we propose a detailed case study in formulating and solving the network coding-enabled optical networks, demonstrating the efficacy of the \textit{optical-computing-enabled network}, and highlighting the unique challenges tied with greater complexities in network design problems, compared to optical-bypass counterpart. 
		
	\end{abstract}
	
	\begin{keyword}
		Optical Communication-Computing Integrated Network \sep Optical-computing-enabled Network \sep In-network Optical Computing \sep Optical-bypass Network  \sep Computed Lightpath \sep Integrated Lightpath \sep Optical-layer Intelligence \sep Optical Aggregation \sep Optical XOR \sep Routing, Wavelength and Network Coding Assignment \sep Wavelength and Computing Assignment \sep Optical Network Design and Planning 2.0 \sep Integer Linear Programming.
	\end{keyword}
	
\end{frontmatter}


\section{Introduction}
By the end of 2023, it has been known that the Internet accessibility has reached to more than two-thirds of the global population and the remaining will soon be connected thanks to excellent initiatives aiming at connecting the unconnected and thus closing the connectivity gap \cite{Cisco20}. As broadband Internet services have and will become ubiquitous globally, network operators both at global, regional and national scales thus need to ensure the availability of adequate resources to sustain the quality and scale of future connectivity, particularly towards faster, more transparent, and greener communication networks \cite{20years, goptics1, ir4}. Indeed, the majority of Internet traffic today has been carried over optical communications and networks infrastructure, spanning from access, metro to core ones, providing high-capacity channels that make it possible for a plethora of services nowadays, such as Internet-of-things (IoT), Big Data and Artificial Intelligence (AI). In practice, the current fiber-optic communication is operated on a particular wavelength region called C-band featuring minimal and nearly uniform transmission losses across the band. Although C-band was often viewed as an infinite capacity in the context of traditional voiced-dominated services, the fact that (i) its bandwidth is physically finite (i.e. about 5 THz), and (ii) there is an explosive increase in terms of traffic driven by data-centric and bandwidth-intensive applications, could lead to its exhaustion as resource. This well-known phenomenon is often referred as the capacity crunch problem \cite{futureoptics1, futureoptics2}. On the other hand, as coherent transmission has been the mainstream in optical communication systems and digital signal processing play a critical role in the coherent transceiver on the transmit and receive sides, continuing to increase the system capacity will be eventually be bottle-necked by the electronic processing limitation. This phenomenon is known as the end of Moore’s law. Sustaining the explosive Internet traffic growth driven by the future connectivity requirements will thus entail ground-breaking innovations in the realm of optical communication and networking. Indeed, optical transport networks have been advancing year-by-year thanks to unabated technological and architectural improvements. In the next 10-year time-frame, major changes encompassing both technological and architectural aspects will be envisioned, foreseeing a disruptive capacity expansion to support massive connectivity of a hyper-connected world in a greater cost and energy efficiency. On one hand, technology-based approaches relying on the development of more advanced transmission systems, higher-order modulation formats and recently wideband transmission, pave the way for expanding the system capacity by orders of magnitude \cite{NICT}. Particularly, the two emerging Multi-band (MB) and Space-Division Multiplexing (SDM) transmission technologies have been garnering significant attention from both the academia and leading industrial players to boost the system capacity by many orders-of-magnitude. For SDM technology, the main principle is to exploit the parallelization in transmission over a bundle of multiple fibers and/or over multi-core/mode fibers (MCF/MMF). Although SDM technology holds the promise of more than hundred-fold improvements in fiber capacity which makes it a rewarding candidate for supporting future connectivity, the main drawback lies in the requirement of additional new types of fibers and optical components such as wavelength-selective switches, filters, and amplifiers which incurs a complex infrastructure upgrade and thus raises commercial concerns with the current stages of technologies and implementation \cite{gsnr, trend1, trend2, trend3}. As an alternative potential capacity-enhancement solution, the MB-based transmission  exploiting the remaining available spectrum in the conventional single-mode fibers (SMFs) in addition to the C-band represents a viable route to sustain the traffic growth in the near-term perspective. As conventional networks likely reaching their physical limit, a.k.a, capacity crunch, the MB transmission may appear as a practical and promising upgrade direction with minimal disruption with the current infrastructure. Indeed, it has been experimentally reported that a more than ten-fold capacity expansion could be achieved with the MB transmission \cite{thesis2024, gsnr}. \\ 

On the architectural front where the major goal is less on enlarging the system capacity, but more on reducing the effective network traffic and thus dropping the capital and operational cost of per-transmitted bits, optical networking has been gradually migrated from the simple point-to-point connections based on the optical-electrical-optical (O-E-O) mode to optical-bypass operations; the key difference being that the transiting lightpaths in optical-bypass mode, rather than undergoing unnecessary and costly O-E-O conversions, could profit from optical cross-connect at intermediate nodes en route from the source to destination \cite{efficient, Simmons}. Thanks to massive gains enabled by optical-bypass mode and rapid advances in both devices, systems and networking technologies permitting the wide deployments of optical-bypass-enabled networks, such networks have remained the architecture of choice by worldwide operators in the last two decades since the year 2000s \cite{all-optical}. In optical-bypass framework, the add/drop and cross-connect functions constitute the fundamental operations in handling the traffic at the optical layer, where the underlying principle lies in the fact that in cross-connecting in-transit lightpaths over an intermediate node, these lightpaths must be guarded from each other in either time, frequency or spatial domain, to avoid interference which is treated as destructive. \cite{nodearchitecture, all-optical}. This turns out to be a fundamental limitation as various optical computing operations could be performed between such transitional lightpaths to produce the new ones which could be spectrally more-efficient than its inputs and/or could serve special computing purposes at scale such as training large-scale AI models. Inspired by the rapid progresses in the realm of optical computing enabling the controlled interference of optical channels that are tailored to various computing capabilities, we envision a different perspective to turn around the long-established wisdom in optical-bypass network, by putting the optical channel interference to a good use, resulting into the so-called optical-computing-enabled network and that potentially marks the next frontier of optical network architecture with the arrival of optical communication-computing integrated networks. Our proposal is essentially defined by the added capability of optical nodes leveraging the superposition of transitional lightpaths to compute new ones of greater capacity efficiency and/or for special computing purposes such as training large-scale AI models. In particular, two illustrative examples underlining the potential merits of bringing about in-network optical computing functions, that is, optical aggregation and optical XOR gate, are presented. The new optical computing capabilities armed at optical nodes therefore demand for a radical change in the manner that network problems should be formulated and their associated algorithmic solutions be investigated. This paradigm shift is collectively referred as optical network design and planning 2.0. In this paper, we propose a case study for network-coding-enabled optical networks, showing the efficacy of optical-computing-enabled network and the unique challenges that are tied with greater complexities in network design problems compared to the optical-bypass counterpart \cite{aina, hai_tnsm, hai_oft24}. \\

The remainder of the paper is structured as follows. In Section 2, the concept of optical-computing-enabled paradigm is described in-depth, and the applications of two
optical computing operations, i.e., optical aggregation and optical XOR, are highlighted. The computational impact and intricacies for network design and planning in the paradigm of optical-computing networking, is also addressed. In order to reveal the more complicated network design problem arisen in the optical-computing-enabled network, a mathematical formulation for solving the routing, wavelength and network coding assignment problem, which is based on the integer linear programming (ILP) model, is presented in Section 3. In Section 4, we show some numerical evaluations, where our proposal which leverages the use of optical XOR encoding within the framework of optical-computing enabled mode, is compared against the traditional optical-bypass networking paradigm, using realistic COST239 and NSFNET network topologies. We also highlight the critical difference between solving the traditional routing and wavelength assignment (RWA) problem arisen in the context of optical-bypass mode to its evolved variant, i.e. the routing, wavelength and network coding assignment problem (RWNCA) newly appeared in the optical-computing-enabled network. Finally, Section 5 concludes the paper and highlights potential future works.

\label{intro}

\section{Optical Computing-Communication Integrated Network}
In 1965, a groundbreaking prediction appeared in the Electronics magazine which has then become the famous Moore's law that governed the evolution of integrated circuits complexity for over five decades, enabling the use of billions of transistors in chip designs. However, the era of predictable transistor scaling is showing signs of abating; new paradigm shifts are therefore constantly sought out to advance the computing performance, particularly in the context of massive computing requirements from AI-driven applications. One promising solution is the transition to photonic-based computing and in that endeavor, silicon photonics demonstrated by Intel in 2013, showcased reduced power consumption and size while achieving superior speed compared to the conventional electronic computing counterparts. Leading tech companies have then been actively exploring silicon photonics such as Broadcom developing 25.6 Tb/s and 51.2 Tb/s co-packaged switches, integrating PICs with ASIC SerDes. Cisco has been developing large-scale silicon photonic PICs for diverse applications while HP has been using silicon photonics for HPC applications, achieving high bandwidth and low-power operation. IBM has recently announced a $\$3$ billion investment to explore next-generation low-power transistors and silicon photonics, among other pioneering initiatives \cite{nature, nature2, nature3}. In the wake of AI-driven computing, massive investments have also been poured to light-based solutions such as LightIntelligence and LightMatter companies to accelerate the AI developments at the speed of light \cite{photonicmit1, photonicmit2, optical_processing_5, xor3}. The rapid rise of optical computing thus begs a question of a new network architecture that could offer fertile soil for successful utilization of photons in computing. \\

For many years, optical communication and networks have been serving the transportation of information from one point to another while the computing is performed at the electrical layer. Optical layer thus plays a static role in handling the traffic, where en route from the source to destination, the optical channel is simply cross-connected over intermediate nodes without further processing / computing operations. Inspired by the renewed interests and then rapid advances in optical computing technologies, it would be envisioned to have a paradigm shift from the current optical communication networks to the optical communication-computing integrated networks, where both the transmission and computing functions are available at the optical layer. Our proposal is named as optical-computing-enabled network to differentiate itself from the currently used optical-bypass mode. As a major departure from the legacy optical-bypass, optical-computing-enabled network featuring the interaction of transitional optical channels paves the way for redefining the optical network architecture, turning around the conventional assumption of keeping the transitional lightpaths untouched. In this context, the optical-computing-enabled framework could be foreseen as a part of the next evolution of optical-bypass networking. \\

This section is dedicated to illustrate the efficient use and consequently network-wide impact of introducing two optical computing operations, namely, optical aggregation / de-aggregation and optical XOR into the optical layer of the optical transport networks. It is worth noting that the enabling technologies for realizing such these two optical computing operations have been rapidly accelerating, paving the way for technological readiness of upgrading optical nodes with optical computing functionalities. Besides, while the discussion in this section is restricted to the two operations, it does not exclude other optical computing operations that could be performed at the lightpath scale. Indeed, as photonic computing technologies move forward, a wide range of computing functions could be technologically feasible and such advances will be expected to have massive impacts to optical networks from both the design, planning, operation and management standpoints.  \\

\subsection{Optical Aggregation / De-aggregation}
It remains an essential function in the operation of optical networks concerning the efficient aggregation of lower-speed channels into a single higher-speed one so that high-capacity optical channels could be optimally utilized. The more efficient the aggregation is, the greater capacity efficiency could be achieved thanks to freeing up the lightpaths of lower wavelength utilization. In optical transport networks, the aggregation functionality has been conventionally performed in the electronic domain which includes terminating the optical channels, re-assembling, re-modulating and eventually back-converting them to the optical domain. As the optical channels operate at an increasingly higher rate, that traditional way of aggregation poses many limitations and clearly it is not scalable for the era of very high bit-rate operations. As a potential solution, the concept of optical aggregation has been recently proposed, investigated and experimentally demonstrated \cite{agg11, agg12}. The leading tech company for this revolutionary effort includes INFINERA, stepping up to develop a new ecosystem of devices and components with the capability of transforming the traditional operation of optical nodes \cite{p2mp0, p2mp1}. \\

From the implementation perspective, optical aggregation and de-aggregation have been realized by exploiting the nonlinear effects when two or more optical channels co-propagate in a nonlinear medium. Specifically, the second and/or third-order susceptibility of nonlinear mediums such as highly nonlinear fiber and semiconductor optical amplifiers resulting in nonlinear phenomenons, i.e., four-wave mixing, cross-phase modulation, self-phase modulation, and cross-gain modulation have been the major mechanism to implement optical aggregation and de-aggregation functions \cite{agg3, agg4, agg5, agg6}. From the functional perspective, two or more optical channels of lower bit-rate and lower-order modulation format could be optically added together into a single higher bit-rate and higher-order modulation format thanks to using an optical aggregator. In the following illustrative case, the utilization of an optical aggregator to combine two QPSK signals into a single 16-QAM channel and an optical de-aggregator for the vice-versa are examined. Figure 1 depicts the schematic diagram for the addition of two QPSK channels of lower bit-rate into a single 16-QAM channel of higher rate, leading to two-fold improvement in the spectral efficiency. \\

\begin{figure}[!ht]
	\centering
	\includegraphics[width=0.7\linewidth, height = 7cm]{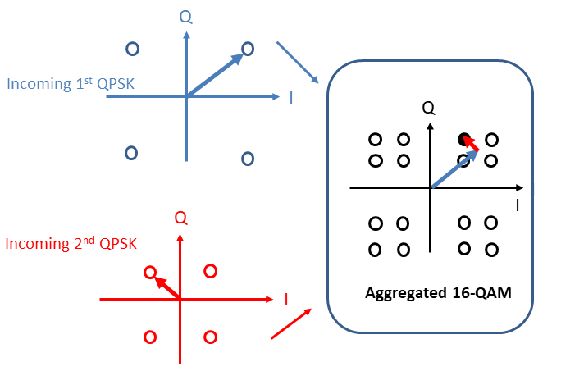}
	\caption{Schematic representation of the optical aggregation operation: Adding two QPSK signals to generate the resultant one 16-QAM signal}
	\label{fig:i1}
\end{figure}

In revealing how to achieve network-wide profit from using the above optical aggregator, we first consider the conventional way of accommodating the traffic demands in optical-bypass networking. Figure 2 shows the routing and wavelength assignment for two demands $a$ and $b$ of the same line-rate 100G and format QPSK. In the absence of a wavelength converter, it is well-known that due to the wavelength uniqueness constraint, two wavelengths are required on link $XI$ and $IC$. Now lets switch to a different perspective, supposing that at node $X$, the optical aggregation is enabled. Under such new operational paradigm, the two 100G QPSK transitional lightpaths $a_{\lambda_1}$ and $b_{\lambda_1}$ crossing the same node $X$ could be optically interfered with each other to generate the output signal of 200G which is modulated on 16-QAM format and on the same wavelength $\lambda_1$ (i.e., $(a+b)_{\lambda_1}$) and that aggregated lightpath carrying the traffic of both demand $a$ and $b$ is routed all the way to the destination node. At the common destination node $C$, the aggregated lightpath undergoes the de-aggregation process to extract constituent ones and such decomposition operation could be performed either in the optical and electrical domain. It should be bear in mind that the computing sense in this context is interpreted as the addition of bits-per-symbol, that is, 2 bits/symbol for QPSK and 4 bits/symbol for 16-QAM. In comparing two approaches for accommodating the traffic demands, it is clearly observed from the Fig. 3 that the optical-computing-enabled one results into greater capacity efficiency as for the whole network, a single wavelength is required compared to the two wavelengths requirement for the optical-bypass one.  \\

\begin{figure}[!ht]
	\centering
	\includegraphics[width=0.6\linewidth, height = 8cm]{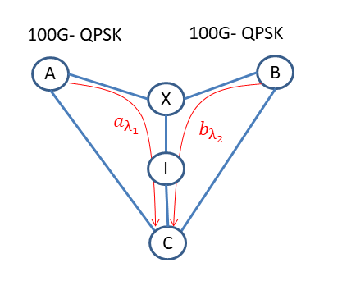}
	\caption{Provisioning two requests in the optical-bypass networking}
	\label{fig:i2}
\end{figure}

\begin{figure}[!ht]
	\centering
	\includegraphics[width=\linewidth, height = 7cm]{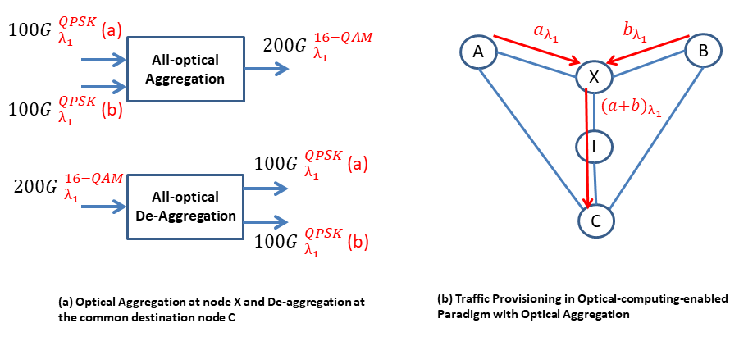}
	\caption{Optical-computing-enabled paradigm with optical aggregation and de-aggregation}
	\label{fig:i3}
\end{figure}

It is worth noting that the optical-computing-enabled paradigm paves the way for a new dimension, that is, the interference of transitional lightpaths for computing purposes and it leads to new network design and planning algorithms so that network-wide gain could be fully attained. In the case of the above optical aggregation, the added complexity lies in selecting pairs of lightpaths for aggregation, the corresponding aggregation node, and more importantly, the determination of the route and wavelength for the aggregated lightpaths.  \\

\subsection{Optical XOR Encoding / Decoding}
Photonic-based logic gates technologies have been rapidly accelerating in recent years, making it feasible to perform the bit-wise exclusive-or (XOR) between optical signals of very high bit-rates and/or different modulation formats \cite{xor3, nc_others10}. A schematic representation of such device is depicted in Fig. 5(a), where two optical signals carrying 100G modulated on the same wavelength and format QPSK are optically XOR-coded to produce the output X of the same bit-rate, format and wavelength. \\

\begin{figure}[!ht]
	\centering
	\includegraphics[width=0.8\linewidth, height = 5cm]{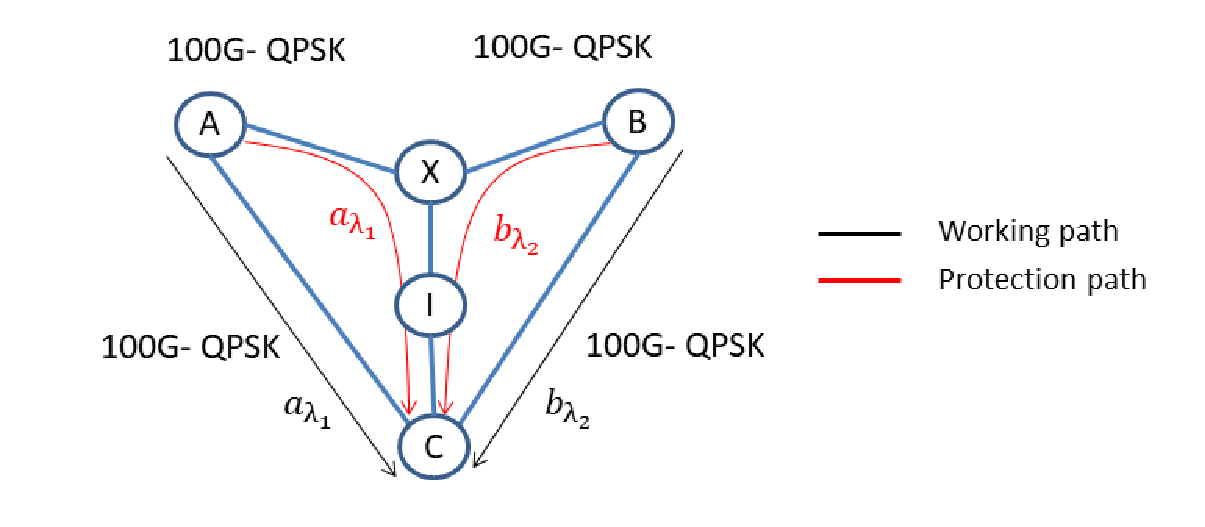}
	\caption{Provisioning two requests with dedicated protection in optical-bypass networking}
	\label{fig:i4}
\end{figure}

\begin{figure}[!ht]
	\centering
	\includegraphics[width=\linewidth, height = 7cm]{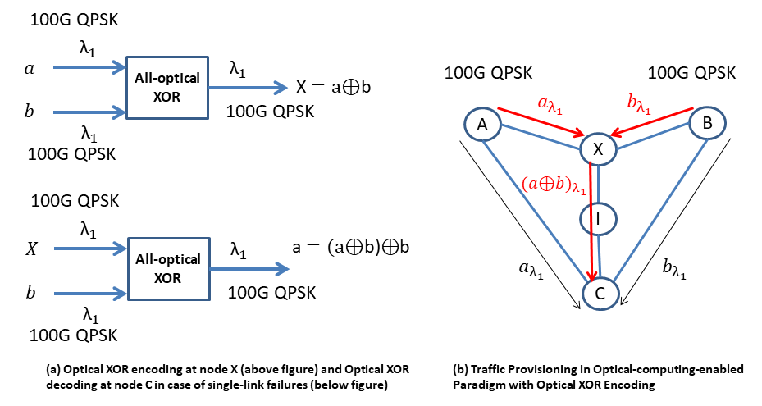}
	\caption{Optical-computing-enabled paradigm with optical XOR encoding and decoding}
	\label{fig:i5}
\end{figure}

The utilization of optical XOR at the optical layer in optical transport networks is illustrated in this part through the protection scenario. Assuming that there are two demands $a$ and $b$ requesting the same bit-rate of 100G from node A and node B, respectively to node C. Both demands are also under the dedicated protection scheme. One way of provisioning such two demands are shown in Fig. 4 for the optical-bypass framework where the route and wavelength for both working and backup lightpaths of both demands are determined subject to typical constraints of wavelength uniqueness and link-disjointedness. It is noticed that as the backup lightpath of demands $a$ and $b$ crossing the same links $XI$ and $IC$, it therefore requires two wavelengths on those links as a consequence of wavelength uniqueness constraint. We now turn the attention to the case of optical-computing-enabled paradigm when node $X$ is armed with the optical XOR encoding capability as shown in Fig. 5(b). Under this assumption, the backup signal of demand $a$ and demand $b$ could thus be optically XOR-encoded with each other at node X to produce the new lightpath $X = a \oplus b$ of the same bit-rate, format and same wavelength as the inputs. That new computed lightpath carrying the encoded traffic between demand $a$ and $b$ is routed all the way from node $X$ to the shared destination node $C$, consuming a single wavelength channel on both link $XI$ and $IC$ which results in a spectral saving of $50\%$ compared to the optical-bypass mode. In facing single-link failure events on the working paths, the recovery capability for both demand $a$ and demand $b$ is guaranteed by making use of the XOR operation on the two remaining signals. Specifically, if the working signal of demand $b$ is lost, it can be retrieved alternatively by the following operation: $b = (a \oplus b) \oplus a$ and the same principle is applied in the lost of the the working signal of demand $a$. \\

As shown in the above illustration, the use of optical XOR encoding in the context of dedicated protection appears to be a good match as it can, on one hand, attain greater capacity efficiency while on the other hand, still retaining the merit of near-immediate recovery speed. In a general case, exploiting the network coding benefits rely on the capability to solve more complicated network design problems. Specifically, the added complexity lies in determining a set of pairs of demand for encoding, together with encoding nodes and the selection of routes and/or transmission parameters for encoded lightpaths. \\

\subsection{Network Design and Planning: Optical-computing-enabled vs. Optical-Bypass Framework}
It is worth noting that the \textit{optical-computing-enabled} paradigm featuring the interference among two or more favorable lightpahts for computing purposes introduces more networking flexibility as a new dimension accounting for the interaction of lightpath is created. Such new dimension clearly poses significant ramifications in both the formulation and solving network design and planning problems
to tap into the potential benefits \cite{hai_comletter, hai_systems, hai_comcom, hai_comcom2, hai_springer5, hai_apnet, hai_ro}.  \\

In optical-bypass networking, the routing and wavelength assignment (RWA) problem remains the central one in design, planning and operation of a network, determining the network efficiency. In essence, solving the traditional RWA problem involves the selection of a route and the assignment of transmission parameters including wavelength/spectrum and/or format for each individual demand subject to a set of typical constraints including mainly the wavelength uniqueness, wavelength continuity and/or contiguity. Unlike in optical-bypass, more complicated network design problems arise in the optical-computing-enabled paradigm due to the interaction of the transitional lightpaths. Specifically, beyond identifying the route and wavelength/spectrum for each demand, the determination of pair of lightpaths to compute and the corresponding node needs to be taken into account. Furthermore, the arrival of special lightpaths resulting from the interaction of two or many demands gives rise to the issue of selecting their route and assigning their wavelength/spectrum. That said, the RWA problem is extended by an additional dimension, which may generally be referred as the computing assignment involving the pairing of in-transit lightpaths for computing purposes. This represents a major departure in the network design and planning, leading to a revisit of the traditional set of algorithms that have been well-developed for optical-bypass networking in many years. In addressing this fundamental shift, a new framework that may collectively be named as optical network design and planning 2.0 should be investigated and developed, encompassing new problems emerging from the various ways that transitional lightpaths could be optically mixed, and the associated  algorithms that include exact/heuristic solutions for solving them could be derived. \\

In the next part, we showcase the problem, entitled, the routing, wavelength and network coding assignment problem (RWNCA) arisen in the utilization of optical XOR in the context of optical-computing-enabled network, along with the mathematical formulation in the form of the integer linear programming model for optimally solving it. 

\section{A Mathematical Formulation for Optical-computing-enabled Network Design with Optical XOR Encoding and Decoding}

This section is focused on the case of designing network-coding-enabled optical networks in supporting a given set of traffic demands such that the wavelength link utilization is minimized. The optical network coding scheme utilized is the simple XOR operation. The optical XOR gate receives input signals of the same wavelength, line-rate and format and then produces the XOR-coded version output of the same wavelength and format and line-rate as inputs. Such XOR coding between signals of the same wavelength features the distinct advantage, that is, the elimination of a probe signal and therefore, could lead to greater cost-efficiency and less operational complexity \cite{xor-model}. Besides, for the practical purpose of easing the operation as the optical XOR is introduced at the optical layer, the optical encoding is permitted to be performed only on the backup signals of demands that share the destination node. We also assume that there is a maxium of one encoding operation for each demand and thus, the decoding, if any, is only allowed at the destination. As a consequence of the aforementioned assumptions, following constraints on the network coding assignment for any two code-able demands must be ensured: i) two demands must have a common destination ii) two demands must use same wavelength iii) the link-disjointedness constraint between two demands' working paths and one’s working to the another’s backup path iv) the two demand's backup paths must have a common sub-path whose one end is the common destination of two demands. \\

\noindent{Inputs:}
\begin{footnotesize}
	\begin{itemize}
		\item $G(V,E)$: A graph models the physical network topology consisting of $|V|$ nodes and $|E|$ fiber links. The beginning and ending node making up a link $e \in E$ is represented by $s(e)$ and $r(e)$, respectively. 
		\item $D$: A set of traffic demands, indexed by $d$. The source and destination node of a demand $d \in D$ are notated respectively as $s(d)$ and $r(d)$, and all demands are assumed to request the same \textit{wavelength capacity} (e.g., 400G)
		\item $W$: A set represents available wavelengths on each fiber link, indexed by $w$. The link capacity measured in number of wavelength is $|W|$ \\
	\end{itemize}
\end{footnotesize}

\noindent{Design Variables:}
\begin{footnotesize}
	\begin{itemize}
		\item $\alpha_{e, w}^{d} \in \{0,1\} $: equals 1 if link $e$ and wavelength $w$ is used for working path of demand $d$, 0 otherwise.
		
		\item $\beta_{e, w}^{d} \in \{0,1\} $: equals 1 if link $e$ and wavelength $w$ is used for the backup path of demand $d$, 0 otherwise.
		
		\item $\theta_{w}^{d} \in \{0,1\} $: equals 1 if wavelength $w$ is used for demand $d$, 0 otherwise.
		
		\item $z_{e, w}^{d, v} \in \{0,1\} $: equals 1 if demand $d$ on wavelength $w$ is encoded at node $v$ and the coding path includes link $e$, 0 otherwise
		
		\item $\delta_{v}^{d} \in \{0, 1\} $: equals 1 if at node $v$, demand $d$ is encoded with another demand, 0 otherwise
		
		\item $f_{d_1}^{d_2} \in \{0, 1\} $: equals 1 if demand $d_1$ is encoded with demand $d_2$, 0 otherwise
		
		\item $\gamma_{e,w}$: equals 1 if wavelength $w$ is used on link $e$, 0 otherwise \\
	\end{itemize}
\end{footnotesize}

\noindent{Objective function:}
\begin{footnotesize}
	\begin{equation} \label{eq:obj}
		\textit{Minimize} \; \sum_{e \in E} \sum_{w \in W}  \gamma_{e,w}
	\end{equation}
\end{footnotesize}

\noindent{Subject to:}
\begin{footnotesize}
	\begin{equation}\label{eq:c1}
		\sum_{w \in W} {\theta^d_w} = 1 \; \; \forall d \in D 
	\end{equation}
	
	\begin{equation} \label{eq:c2}
		\begin{split}
			\sum_{e \in {E}: v\equiv s(e)} {\alpha_{e, w}^{d} (\beta_{e, w}^{d})}-\sum_{e \in {E}: v \equiv r(e)} {\alpha_{e, w}^{d} (\beta_{e, w}^{d})}= \\		
			\begin{cases} 
				\theta_{w}^{d} &\mbox{if } v \equiv s(d) \\ 
				-\theta_{w}^{d}& \mbox{if } v \equiv r(d)\\
				$0$ & otherwise \\
			\end{cases}     \qquad \qquad \forall v \in V, \forall d \in D, \forall w \in W \hfill
		\end{split}
	\end{equation}

	\begin{align} \label{eq:c3} {
			\alpha_{e, w}^{d} + \beta_{e, w}^{d} \leq 1 \qquad \forall d \in D, \forall w \in W, \forall e \in E 
		}
	\end{align}
	
	\begin{align} \label{eq:c4} 
		\begin{split}
			\sum_{d \in D} \alpha_{e, w}^{d} + \sum_{d \in D} \beta_{e, w}^{d} -\frac{1}{2} \sum_{d \in D} \sum_{v \in V} {z_{e, w}^{d, v}} = \gamma_{e,w} \\
			\qquad \forall e \in E, \forall w \in W
		\end{split}
	\end{align}
	
	
	\begin{align} \label{eq:c6} {
			\sum_{v \in V} \delta_{v}^{d} \leq 1 \qquad and \qquad \delta_{v}^{d} = 0 \qquad \mbox{if } v \equiv r(d) \qquad \forall  d \in D
		}
	\end{align}
	
	\begin{align} \label{eq:c7} {
			\sum_{d_2 \in D} f_{d_1}^{d_2} \leq 1 \qquad \forall d_1 \in D
		}
	\end{align}
	
	\begin{equation} \label{eq:c8}
		{f^{d_1}_{d_1}} + \sum_{d_2 \in D: r(d_2) \neq r(d_1)}  {f^{d_1}_{d_2}} = 0 \qquad \forall d_1 \in D
	\end{equation}

	\begin{align} \label{eq:c9} {
			f_{d_1}^{d_2} = f_{d_2}^{d_1} \qquad \forall d_1, d_2 \in D
		}
	\end{align}

	\begin{align} \label{eq:c10} {
			\sum_{d_2 \in D} f_{d_1}^{d_2} = \sum_{v \in V} \delta_{v}^{d_1} \qquad \forall d_1 \in D
		}
	\end{align}
	
	\begin{align} \label{eq:c11} {
			\sum_{w \in W} \sum_{v \in V} z_{e, w}^{d_1, v} \leq \sum_{d_2 \in D} f_{d_1}^{d_2} \qquad \forall d_1 \in D, \forall e \in E
		}
	\end{align}
	
	\begin{align} \label{eq:c12} {
			\sum_{w \in W} z_{e, w}^{d, v}  \leq \delta_{v}^{d}   \qquad \forall d \in D, \forall v \in V, \forall e \in E
		}
	\end{align}

	\begin{align} \label{eq:c13} {
			\sum_{w \in W} \alpha_{e, w}^{d_1} + \sum_{w \in W} \alpha_{e, w}^{d_2} + f_{d_1}^{d_2} \leq 2 \qquad \forall d_1, d_2 \in D, \forall e \in E
		}
	\end{align}
	
	\begin{align} \label{eq:c14} {
			\sum_{w \in W} \alpha_{e, w}^{d_1} + \sum_{w \in W} \beta_{e, w}^{d_2} + f_{d_1}^{d_2} \leq 2 \qquad \forall d_1, d_2 \in D, \forall e \in E
		}
	\end{align}
	
	\begin{align} \label{eq:c15} {
			\theta_{w}^{d_1} - \theta_{w}^{d_2}+f_{d_1}^{d_2} \leq 1 \qquad \forall d_1, d_2 \in D, \forall w \in W
		}
	\end{align}
	
	\begin{align} \label{eq:c16} {
			\theta_{w}^{d_2} - \theta_{w}^{d_1}+f_{d_1}^{d_2} \leq 1 \qquad \forall d_1, d_2 \in D, \forall w \in W
		}
	\end{align}
	
	\begin{align} \label{eq:c17} {
			\delta_{v}^{d_1} - \delta_{v}^{d_2}+f_{d_1}^{d_2} \leq 1 \qquad \forall d_1, d_2 \in D, \forall v \in V
		}
	\end{align}
	
	\begin{align} \label{eq:c18} {
			\delta_{v}^{d_2} - \delta_{v}^{d_1}+f_{d_1}^{d_2} \leq 1 \qquad \forall d_1, d_2 \in D, \forall v \in V
		}
	\end{align}
	
	\begin{align} \label{eq:c19} {
			z_{e, w}^{d, v}  \leq \beta_{e, w}^{d}  \qquad \forall d \in D, \forall v \in V, \forall e \in E, \forall w \in W
		}
	\end{align}
	
	\begin{equation} \label{eq:c20}
		\begin{split}
			\sum_{w \in {W}} (\sum_{e \in E: i=s(e)} z_{e, w}^{d, v} - \sum_{e \in E: i=r(e)} z_{e, w}^{d, v})=\\
			\begin{cases} 
				\delta_{v}^{d} &\mbox{if } i \equiv v \\ 
				-\delta_{v}^{d} & \mbox{if } i \equiv r(d)\\
				0 & \mbox{otherwise}
			\end{cases} \qquad \qquad \forall d \in D, \forall v \in V, \forall i \in V \hfill
		\end{split}
	\end{equation}
	
\end{footnotesize}

The objective function in Equation (1) aims at minimizing the wavelength link cost. The constraints in Equation (\ref{eq:c1}) is to guarantee that all demands are served by finding the proper wavelength. The conservation for both the working and backup flow are ensured by the constraints formulated in Equation (\ref{eq:c2}). The link-disjointedness condition between the working and backup route is captured in Equation (\ref{eq:c3}). The wavelength uniqueness condition on each link is guaranteed by Equation (\ref{eq:c4}). The assumption that each demand has at most one coding node which is different from its destination is ensured by Equation (\ref{eq:c6}). The condition that each demand is coded with at most one another demand of the same destination is indicated by the constraints in Equation (\ref{eq:c7}), Equation (\ref{eq:c8}) and Equation (\ref{eq:c9}). Constraints formulated in Equation (\ref{eq:c10}), Equation (\ref{eq:c11}) and Equation (\ref{eq:c12}) are for coherence purpose, i.e., if a demand is encoded, the respective coding node, coding links and coding wavelength must be found. Constraints given in Equation (\ref{eq:c13}) and Equation (\ref{eq:c14}) are to guarantee that if two demands are encoded with each other, their working routes must be link-disjointed and the working route of one demand must also be link-disjointed with the backup route of another demand. Such constraint are to ensure the recovery capability against any single link failure. The same wavelength condition for code-able demands is captured in Equation (\ref{eq:c15}) and Equation (\ref{eq:c16}). Constraints in Equation (\ref{eq:c17}) and Equation (\ref{eq:c18}) mean that if two demands are encoded together, the same coding node must prevail. The coherence between coding link(s) and backup route is expressed by Equation (\ref{eq:c19}). The last constraint in Equation (\ref{eq:c20}) is the traditional flow conservation one. \\

The above formulation is in the form of an integer linear programming model whose complexity is well-known to be NP-hard. It is worth noting that in addition to the standard variables and constraints representing the route selection and wavelength assignment for each demand, new variables and constraints accounting for the interference of transitional lightpaths for computing purposes have been introduced. Specifically, the existence of variable $z_{e, w}^{d, v} \in \{0,1\}$ and constraints in Equation (\ref{eq:c20}) give the rise to the model one order of magnitude computationally harder than its counterpart, that is, the traditional routing and wavelength assignment in optical-bypass networking. In acknowledging the NP-hard nature of the model, we therefore propose the following scalable heuristic as described in Fig. 6 to be used in large-scale networks .  \\

\begin{figure}[!ht]
	\centering
	\includegraphics[width=0.8\linewidth, height = 10cm]{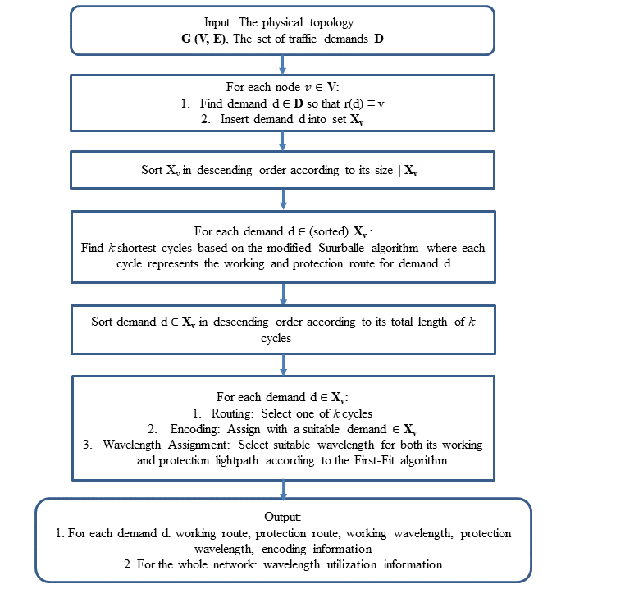}
	\caption{Flowchart of the heuristic algorithm for solving the Routing, Wavelength and Computing Assignment Problem}
	\label{fig:flowchart}
\end{figure}

\section{Numerical Simulation Results}
\begin{figure}[!ht]
	\centering
	\includegraphics[width=\linewidth, height = 8cm]{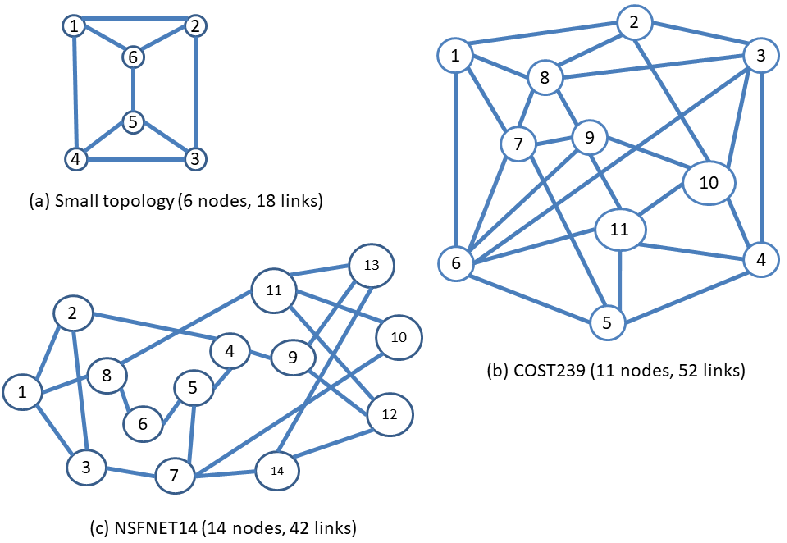}
	\caption{Network Topologies under Investigation}
	\label{fig:topo}
\end{figure}

This section is dedicated to reveal numerical simulation results drawing on a comparative evaluation between our proposal that leverages the efficient use of optical XOR encoding within the framework of optical-computing-enabled networks and the traditional optical-bypass networking. The comparison is experimented on the realistic COST239 and NSFNET network topologies which are shown in Fig. \ref{fig:topo}. The considered performance metric is the conventional wavelength link cost that represents the spectrum utilization efficiency in supporting a given set of traffic demands. Two designs, w-NC and NC, are brought into consideration where the former refers to the design based on the solving the routing and wavelength assignment problem in optical-bypass networking while the latter is obtained from solving the more advanced problem, i.e., routing, wavelength and network coding assignment in optical-computing-enabled networks.  \\

As both the RWA and RWNCA problem is known to be NP-hard complexity, we first evaluate the optimal solutions from solving their ILP models on a small-scale topology including 6 nodes of all degree three, as shown in Fig. 6(a) and compare that with the heuristic ones. The traffic under consideration is generated randomly between nodes with uniform bit-rate requirement (i.e., one wavelength capacity) and the fiber capacity is assumed to be large enough to support all demands (i.e., in our studied case, it is set to be 40 wavelengths). The traffic loads are simulated to represent various conditions from the light, medium and high one corresponding to $30\%$, $70\%$ and $100\%$ (full-mesh) node-pair traffic exchange. Apart from the full-mesh case, for each other traffic condition, there are 20 instances to be simulated. The result in Table 1 is thus averaged across 20 samples. For the network coding-based design (NC) in full-mesh traffic, as the execution time was observed to be exceedingly long and thus, the non-optimal results were collected after 10 hours of running. As revealed in Table 1, the well-studied heuristic for w-NC achieved optimal results which are on a par with its ILP model while the heuristic for NC produced reasonably good solutions with very close gap compared to its ILP model, avoiding the overly long computational time. In view of the sub-optimal nature of heuristic algorithms, the gain obtained by these algorithms were slightly reduced compared to the one from ILP \cite{aina}. \\

\begin{table*}[!ht]
	\caption{Performance comparison between NC-based (NC) and conventional design (w-NC) \cite{aina}}
	\label{tab: r1}
	\centering
	\begin{tabular}{|c|c|c|c|c|c|c|}
		\hline
		\multirow{2}{*}{Load} & \multicolumn{3}{|c|}{Optimal solution from the ILP-based model}  & \multicolumn{3}{|c|}{Heuristic} \\
		\cline{2-7}
		& w-NC & NC & Relative Gain & w-NC & NC & Relative Gain\\ 
		\hline
		$30 \%$ & $32.6$ & $30.6$ & Max: $9\%$, Mean: $6\%$ & $32.6$  & $31.2$ & Max: $9\%$, Mean: $4\%$\\
		\hline
		$70 \%$ & $75.8$ & $67.9$ & Max: $12\%$, Mean: $10\%$ & $75.8$ & $70.6$ & Max: $9\%$, Mean: $7\%$\\
		\hline
		$100\%$ & $108$ & \underline{$99$} & Max=Mean= $8\%$ & 108 & 100 & Max=Mean= $7\%$\\
		\hline
	\end{tabular}
\end{table*}

As the heuristic algorithm's performance for the NC-based design has been verified on the small-scale topology, that algorithm was then used for the larger networks, that is, NSFNET and COST239 topologies, under the same setting about the traffic generation, fiber capacity and number of traffic instances. The obtained results were shown in Table 2. It is observed that up to about $8\%$ gain could be achieved with the NSFNET network. For the more densely connected COST239 network, the lower gain of up to $5\%$ was realized. It is worth noting that in our studied cases, the NC-based design is consistently better than that from the w-NC design, resulting in therefore an improved capacity efficiency. Compared to the findings on the O-E-O case as reported in \cite{icc}, where the gain was known to be up to $20\%$, there was a reduced gain in the all-optical case. This may be due to the wavelength-related constraints for network coding assignments, curbing the coding capability among the demands. Moreover, it should be noted that the gain is highly dependent on the structure of the network topology, traffic and network design algorithms. 

\begin{table*}[!ht]
	\centering
	\caption{Numerical Results for Realistic Topologies \cite{aina}}
	\begin{tabular}{|c|c|c|c|c|c|}
		\hline
		Network & Load & w-NC & NC & Relative Gain & Average No of Coding Operations \\
		\hline
		\multirow{3}{*}{NSFNET} & $30\%$ & 318.3 & 299.5 & Max = $8\%$, Mean = $6\%$ & 9.7 \\
		\cline{2-6}
		& $70\%$ & 730.4 & 682.4 & Max = $8\%$, Mean = $7\%$ & 24.5 \\
		\cline{2-6}
		& $100\%$ & 1048 & 981 & Max = Mean = $6\%$ & 35 \\
		\hline
		\multirow{3}{*}{COST239} & $30\%$ & 126.2 & 123.3 & Max = $3\%$, Mean = $2\%$ & 1.4 \\
		\cline{2-6}
		& $70\%$ & 295.6 & 285.4 & Max = $5\%$, Mean = $3\%$ & 5.1\\
		\cline{2-6}
		& $100\%$ & 420 & 404 & Max = Mean = $4\%$ & 8 \\
		\hline
	\end{tabular}
	
	\label{table:ta}
\end{table*}

\subsection{A closer look on the difference between the general Routing, Wavelength and Network Coding Assignment Problem versus the traditional Routing and Wavelength Assignment}

In this part, we highlight the critical difference between solving the Routing, Wavelength and Network Coding Assignment Problem and the traditional Routing and Wavelength Assignment through an instance of traffic matrix shown in Table \ref{tab: trafficc}. For simplicity and yet without the loss of generality, the traffic matrix is selected so that there is a high opportunity for encoding between the backup lightpaths. Specifically, there are two source nodes, that is, node 5 and node 10, while there are 6 common destination nodes. For the purpose of illustration, we slightly modify the objective function in Equation (1) to become the Equation (21) as followed.

\begin{footnotesize}
	\begin{equation} \label{eq:obj2}
		\textit{Minimize} \; \sum_{w \in W} x_w + \frac{1}{|E||W|+1} \sum_{e \in E} \sum_{w \in W}  \gamma_{e,w}
	\end{equation}
\end{footnotesize}

The new objective formulated in Equation (21) consists of two weighted sub-objectives where the first and prioritized one is to minimize the number of used wavelengths and the secondary goal is to minimize the wavelength link usage. The priority of constituent objectives and consequently the order of optimization is ensured by putting the proper weights as shown in the equation. \\

Table 4 showcases the routing and wavelength assignment for the working and backup paths of all demands. Note that the solution provides a standard information similar to what is obtained when solving the traditional routing and wavelength assignment. Overall, there are six wavelengths needed to support the traffic demands. However, as there are the interaction between backup lightpaths by the optical XOR operation, the determination of which pair of demands are encoded together and at which node such encoding operation occurs should be found and optimized. Also, as the consequence of encoding backup lightpaths and it results in computed lightpaths of greater efficiency than the original ones, the routing and allocating wavelength for such newly appeared lightpaths must be taken into account. In term of complexity, solving the RWNCA problem therefore is one order of magnitude computationally harder than the traditional RWA. Table 5 highlights the added information obtained from solving RWNCA where the pair of demands for encoding, the encoding node, the route as well as the wavelength for computed lightpaths are optimally provided. It is important to note that the spectral gain from exploiting the optical XOR operation comes at the expenses of solving a more difficult network design problem, which is, RWNCA.  

\begin{table*}[!ht]
	\caption{An Instance of Traffic Matrix}
	\label{tab: trafficc}
	\centering
	\begin{tabular}{|c|cccccccccccccc|}
		\hline
		NodeID & 1 & 2 & 3 & 4 & 5 & 6 & 7 & 8 & 9 & 10 & 11 & 12 & 13 & 14 \\
		\hline
		1  & 0 & 0 & 0 & 0 & 0 & 0 & 0 & 0 & 0 & 0 & 0 & 0 & 0 & 0\\
		
		2 & 0 & 0 & 0 & 0 & 0 & 0 & 0 & 0 & 0 & 0 & 0 & 0 & 0 & 0 \\
		
		3 & 0 & 0 & 0 & 0 & 0 & 0 & 0 & 0 & 0 & 0 & 0 & 0 & 0 & 0 \\
		
		4 & 0 & 0 & 0 & 0 & 0 & 0 & 0 & 0 & 0 & 0 & 0 & 0 & 0 & 0 \\
		
		5 & 1 & 1 & 1 & 1 & 0 & 1 & 1 & 0 & 0 & 0 & 0 & 0 & 0 & 0 \\
		
		6 & 0 & 0 & 0 & 0 & 0 & 0 & 0 & 0 & 0 & 0 & 0 & 0 & 0 & 0 \\
		
		7 & 0 & 0 & 0 & 0 & 0 & 0 & 0 & 0 & 0 & 0 & 0 & 0 & 0 & 0\\
		
		8 & 0 & 0 & 0 & 0 & 0 & 0 & 0 & 0 & 0 & 0 & 0 & 0 & 0 & 0\\
		
		9 & 0 & 0 & 0 & 0 & 0 & 0 & 0 & 0 & 0 & 0 & 0 & 0 & 0 & 0 \\
		
		10 & 1 & 1 & 1 & 1 & 0 & 1 & 1 & 0 & 0 & 0 & 0 & 0 & 0 & 0 \\
		
		11 & 0 & 0 & 0 & 0 & 0 & 0 & 0 & 0 & 0 & 0 & 0 & 0 & 0 & 0\\
		
		12 & 0 & 0 & 0 & 0 & 0 & 0 & 0 & 0 & 0 & 0 & 0 & 0 & 0 & 0 \\
		
		13 & 0 & 0 & 0 & 0 & 0 & 0 & 0 & 0 & 0 & 0 & 0 & 0 & 0 & 0 \\
		
		14 & 0 & 0 & 0 & 0 & 0 & 0 & 0 & 0 & 0 & 0 & 0 & 0 & 0 & 0 \\
		\hline 
		
	\end{tabular}
\end{table*}

\begin{table*}[!ht]
	\caption{Routing and Wavelength Allocation Information for Traffic Demands}
	\label{tab: rwa}
	\centering
	\begin{tabular}{ccccc}
		\hline
		Source Node $\rightarrow$ Destination Node & W-path & W-$\lambda$ & B-path & B-$\lambda$ \\
		\hline
		
		5 $\rightarrow$ 1 & (5-4-2-1) & 5 & (5-7-3-1) & 2 \\
		10 $\rightarrow$ 1 & (10-11-8-1) & 5 & (10-7-3-1) & 1 \\
		
		5 $\rightarrow$ 2 & (5-4-2) & 4 & (5-7-3-2) & 6 \\
		10 $\rightarrow$ 2 & (10-11-8-1-2) & 4 & (10-7-3-2) & 6 \\
		
		5 $\rightarrow$ 3 & (5-4-2-3) & 6 & (5-7-3) & 3 \\
		10 $\rightarrow$ 3 & (10-11-8-1-3) & 1 & (10-7-3) & 5 \\
		
		5 $\rightarrow$ 4 & (5-4) & 1 & (5-7-3-2-4) & 5 \\
		10 $\rightarrow$ 4 & (10-11-1-9-4) & 3 & (10-7-3-2-4) & 4 \\
		
		5 $\rightarrow$ 6 & (5-4-2-1-8-6) & 3 & (5-6) & 5 \\
		10 $\rightarrow$ 6 & (10-11-8-6) & 6 & (10-7-5-6) & 3 \\
		
		5 $\rightarrow$ 7 & (5-7) & 1 & (5-6-8-1-3-7) & 6 \\
		10 $\rightarrow$ 7 & (10-7) & 2 & (10-11-8-1-3-7) & 2 \\
		\hline
	\end{tabular}
\end{table*}

\begin{table*}[!ht]
	\caption{Network Coding Assignment Information between Backup Lightpaths}
	\label{tab: computingassign}
	\centering
	\begin{tabular}{cccc}
		\hline
		
		Computed Lightpaths & Computing Node & Route & Wavelength Assignment \\
		(5 $\rightarrow$ 1) $\oplus$ (10 $\rightarrow$ 1) & 7 &  (7-3-1) & 2 \\
		
		(5 $\rightarrow$ 2) $\oplus$ (10 $\rightarrow$ 2) & 7 &  (7-3-2) & 5 \\
		
		(5 $\rightarrow$ 3) $\oplus$ (10 $\rightarrow$ 3) & 7 &  (7-3) & 6 \\
		
		(5 $\rightarrow$ 4) $\oplus$ (10 $\rightarrow$ 4) & 7 &  (7-3-2-4) & 4 \\
		
		(5 $\rightarrow$ 7) $\oplus$ (10 $\rightarrow$ 7) & 8 &  (8-1-3-7) & 1 \\
		
		\hline
	\end{tabular}
\end{table*}

\section{Summary}
In supporting the connectivity of the future driven by the fusion of digital and physical world, network operators have been constantly seeking out innovative solutions to upgrade their network infrastructures so that more traffic could be supported in a greater cost and energy efficiency manner, in addition to supporting stricter requirements on resilience, security and latency. From transmission perspectives, space division multiplexing and ultra-wideband optical technologies have been proposed, actively investigated and progressively experimented to increase the per-fiber capacity by orders-of-magnitude, marking the paradigm shift compared to the current legacy infrastructure. On the parallel front, optical computing has been advancing rapidly to support emerging large-scale AI training services. Optical network architecture has though remained essentially unchanged for the two recent decades since 2000s with the dominance of optical-bypass mode that has successfully integrated technological innovations of its time such as long-haul coherent transmission and reconfigurable optical add/drop multiplexer. This context therefore begs a new architecture that holds the promises of making best uses of technological advances, that is, new optical transmission and optical computing, to support massive connectivity, including future pervasive AI computing traffic. \\

In this paper, we have challenged the status quo with a new perspective of integrating optical computing layer into the traditional legacy optical communication infrastructures, resulting into the new realm of optical communication-computing integrated networks. Our proposal was named as optical-computing-enabled network whose the underlying principle is to reverse the wisdom in optical-bypass operation, that is, instead of keeping in-transit lightpaths over an intermediate node apart from each other in either time, frequency or spatial domain, exploiting the superimposing of such lightpaths in the optical domain for computing purposes is proposed as a way to achieve greater network efficiency. Two illustrative examples highlighting the efficient uses of optical aggregation and optical XOR have been presented and contrasted with the optical-bypass mode. In addressing the new operational paradigm enabled by optical-computing-enabled mode, we have then presented the mathematical formulation for optimal designs of network coding-enabled optical networks. Numerical results evaluating our proposal on the realistic networks, COST239 and NSFNET have been provided, demonstrating its efficacy in comparison with the conventional operation in optical-bypass mode. We have also gone deeper to pinpoint the critical difference in solving the new routing and resource allocation arisen in optical-computing-enabled mode, that is, routing, wavelength and network coding assignment (RWNCA) and the traditional routing and wavelength assignment (RWA). \\

The inherent merit of light over electronic for computing, particularly in the background of ending of Moore's law and massive investment for AI accelerators has put light-based computing solutions at a rapid growth than ever. The rise of optical computing and the quest for both capacity expansion and energy efficiency in optical transport networks have thus been creating a ripe environment for the integration of optical communication and computing infrastructure, laying the foundation for realizing the optical-layer intelligence. This entails a new operational paradigm for optical transport networks, hinting at a radical change in network design problems formulation as well as algorithm developments to tap into new opportunities enabled by the seamless optical computing and communication integration. Various remaining challenges spanning from devices, systems, and networks requiring multidisciplinary efforts will be needed to make optical-computing-enabled mode a practical reality. 


\section*{Conflict of interest}
The authors declare that they have no conflict of interest.

\bibliographystyle{elsarticle-num}

\bibliography{arxiv_IoT}

\begin{thebibliography}{10}
\expandafter\ifx\csname url\endcsname\relax
  \def\url#1{\texttt{#1}}\fi
\expandafter\ifx\csname urlprefix\endcsname\relax\def\urlprefix{URL }\fi
\expandafter\ifx\csname href\endcsname\relax
  \def\href#1#2{#2} \def\path#1{#1}\fi

\bibitem{Cisco20}
Cisco, Cisco annual internet report (2020).

\bibitem{20years}
P.~J. Winzer, D.~T. Neilson, A.~R. Chraplyvy,
  \href{http://www.opticsexpress.org/abstract.cfm?URI=oe-26-18-24190}{Fiber-optic
  transmission and networking: the previous 20 and the next 20 years}, Opt.
  Express 26~(18) (2018) 24190--24239.
\newblock \href {http://dx.doi.org/10.1364/OE.26.024190}
  {\path{doi:10.1364/OE.26.024190}}.
\newline\urlprefix\url{http://www.opticsexpress.org/abstract.cfm?URI=oe-26-18-24190}

\bibitem{goptics1}
A.~Lord, C.~White, A.~Iqbal, Future optical networks in a 10 year time frame,
  in: 2021 Optical Fiber Communications Conference and Exhibition (OFC), 2021,
  pp. 1--3.

\bibitem{ir4}
R.~Sabella, P.~Iovanna, G.~Bottari, F.~Cavaliere,
  \href{http://jocn.osa.org/abstract.cfm?URI=jocn-12-8-264}{Optical transport
  for industry 4.0}, J. Opt. Commun. Netw. 12~(8) (2020) 264--276.
\newblock \href {http://dx.doi.org/10.1364/JOCN.390701}
  {\path{doi:10.1364/JOCN.390701}}.
\newline\urlprefix\url{http://jocn.osa.org/abstract.cfm?URI=jocn-12-8-264}

\bibitem{futureoptics1}
A.~Lord, C.~White, A.~Iqbal, Future optical networks in a 10 year time frame,
  in: 2021 Optical Fiber Communications Conference and Exhibition (OFC), 2021,
  pp. 1--3.

\bibitem{futureoptics2}
P.~J. Winzer, Capacity scaling through spatial parallelism: From subsea cables
  to short-reach optical links, in: 2021 Optical Fiber Communications
  Conference and Exhibition (OFC), 2021, pp. 1--1.

\bibitem{NICT}
NICT, \href{https://www.nict.go.jp/en/press/2021/07/12-1.html}{Demonstration of
  world record: 319 tb/s transmission over 3,001 km with 4-core optical fiber}
  (2021).
\newline\urlprefix\url{https://www.nict.go.jp/en/press/2021/07/12-1.html}

\bibitem{gsnr}
R.~Kalkunte, R.~K. Jana, S.~Ferdousi, A.~Srivastava, A.~Mitra, M.~Tornatore,
  A.~Lord, B.~Mukherjee,
  \href{https://doi.org/10.1007/s11107-024-01023-6}{Gsnr-aware resource
  re-provisioning for c to c+l-bands upgrade in optical backbone networks},
  Photonic Network Communications\href
  {http://dx.doi.org/10.1007/s11107-024-01023-6}
  {\path{doi:10.1007/s11107-024-01023-6}}.
\newline\urlprefix\url{https://doi.org/10.1007/s11107-024-01023-6}

\bibitem{trend1}
P.~J. Winzer, K.~Nakajima, C.~Antonelli, Scaling optical fiber capacities
  [scanning the issue], Proceedings of the IEEE 110~(11) (2022) 1615--1618.
\newblock \href {http://dx.doi.org/10.1109/JPROC.2022.3212229}
  {\path{doi:10.1109/JPROC.2022.3212229}}.

\bibitem{trend2}
P.~J. Winzer,
  \href{https://opg.optica.org/jocn/abstract.cfm?URI=jocn-15-10-783}{The future
  of communications is massively parallel}, J. Opt. Commun. Netw. 15~(10)
  (2023) 783--787.
\newblock \href {http://dx.doi.org/10.1364/JOCN.496992}
  {\path{doi:10.1364/JOCN.496992}}.
\newline\urlprefix\url{https://opg.optica.org/jocn/abstract.cfm?URI=jocn-15-10-783}

\bibitem{trend3}
M.~Shtaif, C.~Antonelli, A.~Mecozzi, V.~X. Chen, {The Information Capacity of
  the Fiber-Optic Channel: Bounds and prospects. }\href
  {http://dx.doi.org/10.1364/opticaopen.24864648.v2}
  {\path{doi:10.1364/opticaopen.24864648.v2}}.

\bibitem{thesis2024}
M.~{van den Hout}, Ultra-wideband and space-division multiplexed optical
  transmission systems, Phd thesis 1 (research tu/e / graduation tu/e),
  Electrical Engineering, proefschrift. (Feb. 2024).
\newblock \href {http://dx.doi.org/10.6100/jnxx-6t19}
  {\path{doi:10.6100/jnxx-6t19}}.

\bibitem{efficient}
A.~Saleh, J.~M. Simmons, Technology and architecture to enable the explosive
  growth of the internet, Communications Magazine, IEEE 49~(1) (2011) 126--132.
\newblock \href {http://dx.doi.org/10.1109/MCOM.2011.5681026}
  {\path{doi:10.1109/MCOM.2011.5681026}}.

\bibitem{Simmons}
J.~M. Simmons, Optical Network Design and Planning, 2nd Edition, Springer
  Publishing Company, Incorporated, 2014.

\bibitem{all-optical}
A.~Saleh, J.~M. Simmons, All-optical networking: Evolution, benefits,
  challenges, and future vision, Proceedings of the IEEE 100~(5) (2012)
  1105--1117.
\newblock \href {http://dx.doi.org/10.1109/JPROC.2011.2182589}
  {\path{doi:10.1109/JPROC.2011.2182589}}.

\bibitem{nodearchitecture}
B.~Collings, M.~Filer,
  \href{https://doi.org/10.1007/978-3-030-16250-4$\_$8}{Optical Node
  Architectures}, Springer International Publishing, Cham, 2020, pp. 259--286.
\newblock \href {http://dx.doi.org/10.1007/978-3-030-16250-4$\_$8}
  {\path{doi:10.1007/978-3-030-16250-4$\_$8}}.
\newline\urlprefix\url{https://doi.org/10.1007/978-3-030-16250-4$\_$8}

\bibitem{aina}
D.~T. Hai, I.~Woungang, On network design and planning 2.0 for
  optical-computing-enabled networks, in: L.~Barolli (Ed.), Advanced
  Information Networking and Applications, Springer Nature Switzerland, Cham,
  2024, pp. 91--102.

\bibitem{hai_tnsm}
D.~T. Hai, On routing, wavelength, network coding assignment, and protection
  configuration problem in optical-processing-enabled networks, IEEE
  Transactions on Network and Service Management 20~(3) (2023) 2504--2514.
\newblock \href {http://dx.doi.org/10.1109/TNSM.2023.3283880}
  {\path{doi:10.1109/TNSM.2023.3283880}}.

\bibitem{hai_oft24}
D.~T. Hai,
  \href{https://www.sciencedirect.com/science/article/pii/S1068520024000750}{What
  comes after optical-bypass network? a study on optical-computing-enabled
  network}, Optical Fiber Technology 84 (2024) 103730.
\newblock \href {http://dx.doi.org/https://doi.org/10.1016/j.yofte.2024.103730}
  {\path{doi:https://doi.org/10.1016/j.yofte.2024.103730}}.
\newline\urlprefix\url{https://www.sciencedirect.com/science/article/pii/S1068520024000750}

\bibitem{nature}
P.~L. McMahon, \href{https://doi.org/10.1038/s42254-023-00645-5}{The physics of
  optical computing}, Nature Reviews Physics 5.
\newblock \href {http://dx.doi.org/10.1038/s42254-023-00645-5}
  {\path{doi:10.1038/s42254-023-00645-5}}.
\newline\urlprefix\url{https://doi.org/10.1038/s42254-023-00645-5}

\bibitem{nature2}
W.~Bogaerts, D.~Pérez, J.~Capmany, D.~A.~B. Miller, J.~Poon, D.~Englund,
  F.~Morichetti, A.~Melloni,
  \href{https://doi.org/10.1038/s41586-020-2764-0}{Programmable photonic
  circuits}, Nature 586.
\newblock \href {http://dx.doi.org/10.1038/s41586-020-2764-0}
  {\path{doi:10.1038/s41586-020-2764-0}}.
\newline\urlprefix\url{https://doi.org/10.1038/s41586-020-2764-0}

\bibitem{nature3}
S.~SeyedinNavadeh, M.~Milanizadeh, F.~Zanetto, G.~Ferrari, M.~Sampietro,
  M.~Sorel, D.~A.~B. Miller, A.~Melloni, F.~Morichetti,
  \href{https://doi.org/10.1038/s41566-023-01330-w}{Determining the optimal
  communication channels of arbitrary optical systems using integrated photonic
  processors}, Nature Photonics 18.
\newblock \href {http://dx.doi.org/10.1038/s41566-023-01330-w}
  {\path{doi:10.1038/s41566-023-01330-w}}.
\newline\urlprefix\url{https://doi.org/10.1038/s41566-023-01330-w}

\bibitem{photonicmit1}
Z.~Zhong, M.~Yang, J.~Lang, C.~Williams, L.~Kronman, A.~Sludds,
  H.~Esfahanizadeh, D.~Englund, M.~Ghobadi,
  \href{https://doi.org/10.1145/3603269.3604821}{Lightning: A reconfigurable
  photonic-electronic smartnic for fast and energy-efficient inference}, in:
  Proceedings of the ACM SIGCOMM 2023 Conference, ACM SIGCOMM '23, Association
  for Computing Machinery, New York, NY, USA, 2023, p. 452–472.
\newblock \href {http://dx.doi.org/10.1145/3603269.3604821}
  {\path{doi:10.1145/3603269.3604821}}.
\newline\urlprefix\url{https://doi.org/10.1145/3603269.3604821}

\bibitem{photonicmit2}
M.~Yang, Z.~Zhong, M.~Ghobadi,
  \href{https://doi.org/10.1145/3626111.3628177}{On-fiber photonic computing},
  in: Proceedings of the 22nd ACM Workshop on Hot Topics in Networks, HotNets
  '23, Association for Computing Machinery, New York, NY, USA, 2023, p.
  263–271.
\newblock \href {http://dx.doi.org/10.1145/3626111.3628177}
  {\path{doi:10.1145/3626111.3628177}}.
\newline\urlprefix\url{https://doi.org/10.1145/3626111.3628177}

\bibitem{optical_processing_5}
P.~Minzioni, C.~Lacava, T.~Tanabe, J.~Dong, X.~Hu, G.~Csaba, W.~Porod,
  G.~Singh, A.~E. Willner, A.~Almaiman, V.~Torres-Company, J.~Schröder, A.~C.
  Peacock, M.~J. Strain, F.~Parmigiani, G.~Contestabile, D.~Marpaung, Z.~Liu,
  J.~E. Bowers, L.~Chang, S.~Fabbri, M.~R. V{\'{a}}zquez, V.~Bharadwaj, S.~M.
  Eaton, P.~Lodahl, X.~Zhang, B.~J. Eggleton, W.~J. Munro, K.~Nemoto, O.~Morin,
  J.~Laurat, J.~Nunn, \href{https://doi.org/10.1088/2040-8986/ab0e66}{Roadmap
  on all-optical processing}, Journal of Optics 21~(6) (2019) 063001.
\newblock \href {http://dx.doi.org/10.1088/2040-8986/ab0e66}
  {\path{doi:10.1088/2040-8986/ab0e66}}.
\newline\urlprefix\url{https://doi.org/10.1088/2040-8986/ab0e66}

\bibitem{xor3}
L.-K. Chen, M.~Li, S.~C. Liew, Breakthroughs in photonics 2014: Optical
  physical-layer network coding, recent developments, and challenges, IEEE
  Photonics Journal 7~(3) (2015) 1--6.
\newblock \href {http://dx.doi.org/10.1109/JPHOT.2015.2418264}
  {\path{doi:10.1109/JPHOT.2015.2418264}}.

\bibitem{agg11}
D.~Welch, A.~Napoli, J.~Bäck, W.~Sande, J.~Pedro, F.~Masoud, C.~Fludger,
  T.~Duthel, H.~Sun, S.~J. Hand, T.-K. Chiang, A.~Chase, A.~Mathur, T.~A.
  Eriksson, M.~Plantare, M.~Olson, S.~Voll, K.-T. Wu, Point-to-multipoint
  optical networks using coherent digital subcarriers, Journal of Lightwave
  Technology 39~(16) (2021) 5232--5247.
\newblock \href {http://dx.doi.org/10.1109/JLT.2021.3097163}
  {\path{doi:10.1109/JLT.2021.3097163}}.

\bibitem{agg12}
J.~Bäck, P.~Wright, J.~Ambrose, A.~Chase, M.~Jary, F.~Masoud, N.~Sugden,
  G.~Wardrop, A.~Napoli, J.~Pedro, M.~A. Iqbal, A.~Lord, D.~Welch, Capex
  savings enabled by point-to-multipoint coherent pluggable optics using
  digital subcarrier multiplexing in metro aggregation networks, in: 2020
  European Conference on Optical Communications (ECOC), 2020, pp. 1--4.
\newblock \href {http://dx.doi.org/10.1109/ECOC48923.2020.9333233}
  {\path{doi:10.1109/ECOC48923.2020.9333233}}.

\bibitem{p2mp0}
D.~Welch, A.~Napoli, J.~Bäck, W.~Sande, J.~Pedro, F.~Masoud, C.~Fludger,
  T.~Duthel, H.~Sun, S.~J. Hand, T.-K. Chiang, A.~Chase, A.~Mathur, T.~A.
  Eriksson, M.~Plantare, M.~Olson, S.~Voll, K.-T. Wu, Point-to-multipoint
  optical networks using coherent digital subcarriers, Journal of Lightwave
  Technology 39~(16) (2021) 5232--5247.
\newblock \href {http://dx.doi.org/10.1109/JLT.2021.3097163}
  {\path{doi:10.1109/JLT.2021.3097163}}.

\bibitem{p2mp1}
Y.~Zhang, Q.~Lv, R.~Li, X.~Tian, Z.~Zhu, Planning of survivable
  wavelength-switched optical networks based on p2mp transceivers, IEEE
  Transactions on Network and Service Management (2023) 1--1\href
  {http://dx.doi.org/10.1109/TNSM.2023.3287302}
  {\path{doi:10.1109/TNSM.2023.3287302}}.

\bibitem{agg3}
Q.~Yang, X.~Wang, Q.~Zhang, X.~Xin, R.~Gao, Y.~Tao, Q.~Tian, F.~Tian, Y.~Wang,
  W.~Zhang, H.~Chang, D.~Chen, J.~Qian, All-optical aggregation scheme based on
  joint modulation, in: 2020 Asia Communications and Photonics Conference (ACP)
  and International Conference on Information Photonics and Optical
  Communications (IPOC), 2020, pp. 1--3.

\bibitem{agg4}
A.~Fallahpour, F.~Alishahi, K.~Zou, Y.~Cao, A.~Almaiman, A.~Kordts, M.~Karpov,
  M.~H.~P. Pfeiffer, K.~Manukyan, H.~Zhou, P.~Liao, C.~Liu, M.~Tur, T.~J.
  Kippenberg, A.~E. Willner, Demonstration of tunable optical aggregation of
  qpsk to 16-qam over optically generated nyquist pulse trains using nonlinear
  wave mixing and a kerr frequency comb, Journal of Lightwave Technology 38~(2)
  (2020) 359--365.
\newblock \href {http://dx.doi.org/10.1109/JLT.2019.2959803}
  {\path{doi:10.1109/JLT.2019.2959803}}.

\bibitem{agg5}
A.~E. Willner, A.~Fallahpour, K.~Zou, F.~Alishahi, H.~Zhou, Optical signal
  processing aided by optical frequency combs, IEEE Journal of Selected Topics
  in Quantum Electronics 27~(2) (2021) 1--16.
\newblock \href {http://dx.doi.org/10.1109/JSTQE.2020.3032554}
  {\path{doi:10.1109/JSTQE.2020.3032554}}.

\bibitem{agg6}
H.~Liu, H.~Wang, Z.~Xing, Y.~Ji,
  \href{http://www.osapublishing.org/oe/abstract.cfm?URI=oe-27-21-30158}{Simultaneous
  all-optical channel aggregation and de-aggregation based on nonlinear effects
  for ook and mpsk formats in elastic optical networking}, Opt. Express 27~(21)
  (2019) 30158--30171.
\newblock \href {http://dx.doi.org/10.1364/OE.27.030158}
  {\path{doi:10.1364/OE.27.030158}}.
\newline\urlprefix\url{http://www.osapublishing.org/oe/abstract.cfm?URI=oe-27-21-30158}

\bibitem{nc_others10}
A.~Kotb, K.~E. Zoiros, C.~Guo, 1 tb/s all-optical xor and and gates using
  quantum-dot semiconductor optical amplifier-based turbo-switched
  mach–zehnder interferometer, Journal of Computational Electronics\href
  {http://dx.doi.org/10.1007/s10825-019-01329-z}
  {\path{doi:10.1007/s10825-019-01329-z}}.

\bibitem{hai_comletter}
D.~T. Hai, Leveraging the survivable all-optical wdm network design with
  network coding assignment, IEEE Communications Letters 21~(10) (2017)
  2190--2193.
\newblock \href {http://dx.doi.org/10.1109/LCOMM.2017.2720661}
  {\path{doi:10.1109/LCOMM.2017.2720661}}.

\bibitem{hai_systems}
D.~T. Hai, L.~H. Chau, N.~T. Hung, A priority-based multiobjective design for
  routing, spectrum, and network coding assignment problem in
  network-coding-enabled elastic optical networks, IEEE Systems Journal 14~(2)
  (2020) 2358--2369.
\newblock \href {http://dx.doi.org/10.1109/JSYST.2019.2938590}
  {\path{doi:10.1109/JSYST.2019.2938590}}.

\bibitem{hai_comcom}
D.~T. Hai, A bi-objective integer linear programming model for the routing and
  network coding assignment problem in wdm optical networks with dedicated
  protection, Computer Communications 133 (2019) 51 -- 58.
\newblock \href
  {http://dx.doi.org/https://doi.org/10.1016/j.comcom.2018.08.006}
  {\path{doi:https://doi.org/10.1016/j.comcom.2018.08.006}}.

\bibitem{hai_comcom2}
D.~T. Hai, On routing, spectrum and network coding assignment problem for
  transparent flex-grid optical networks with dedicated protection, Computer
  Communications\href
  {http://dx.doi.org/https://doi.org/10.1016/j.comcom.2019.08.005}
  {\path{doi:https://doi.org/10.1016/j.comcom.2019.08.005}}.

\bibitem{hai_springer5}
D.~T. Hai, \href{https://doi.org/10.1007/s11082-023-05123-x}{Optical networking
  in future-land: from optical-bypass-enabled to optical-processing-enabled
  paradigm}, Optical and Quantum Electronics 55~(864).
\newblock \href {http://dx.doi.org/10.1007/s11082-023-05123-x}
  {\path{doi:10.1007/s11082-023-05123-x}}.
\newline\urlprefix\url{https://doi.org/10.1007/s11082-023-05123-x}

\bibitem{hai_apnet}
D.~T. Hai, M.~Nguyen, I.~Woungang,
  \href{https://doi.org/10.1145/3663408.3665822}{Optical-computing-enabled
  network: A new dawn for optical-layer intelligence?}, in: Proceedings of the
  8th Asia-Pacific Workshop on Networking, APNet '24, Association for Computing
  Machinery, New York, NY, USA, 2024, p. 215–216.
\newblock \href {http://dx.doi.org/10.1145/3663408.3665822}
  {\path{doi:10.1145/3663408.3665822}}.
\newline\urlprefix\url{https://doi.org/10.1145/3663408.3665822}

\bibitem{hai_ro}
D.~T. Hai,
  \href{https://www.sciencedirect.com/science/article/pii/S2666950123001566}{Optical-computing-enabled
  network: An avant-garde architecture to sustain traffic growth}, Results in
  Optics 13 (2023) 100504.
\newblock \href {http://dx.doi.org/https://doi.org/10.1016/j.rio.2023.100504}
  {\path{doi:https://doi.org/10.1016/j.rio.2023.100504}}.
\newline\urlprefix\url{https://www.sciencedirect.com/science/article/pii/S2666950123001566}

\bibitem{xor-model}
C.~Porzi, et~al., All-optical xor gate by means of a single semiconductor
  optical amplifier without assist probe light, in: LEOS '09. IEEE, 2009, pp.
  617--618.
\newblock \href {http://dx.doi.org/10.1109/LEOS.2009.5343425}
  {\path{doi:10.1109/LEOS.2009.5343425}}.

\bibitem{icc}
H.~Overby, et~al., Cost comparison of 1+1 path protection schemes: A case for
  coding, in: ICC 2012, IEEE, 2012, pp. 3067--3072.
\newblock \href {http://dx.doi.org/10.1109/ICC.2012.6363928}
  {\path{doi:10.1109/ICC.2012.6363928}}.

\end{thebibliography}

\end{document}